\documentclass[apj,twocolumn, twocolappendix]{openjournal}
\makeatletter
\def\@eapj@cap@font{\normalfont}
\makeatother
\usepackage{amsmath}
\usepackage{booktabs}
\usepackage{multirow}
\usepackage{color}
\usepackage{soul}
\usepackage{booktabs} 
\usepackage{xspace}

\usepackage[dvipsnames]{xcolor} 

\usepackage[breaklinks,colorlinks,citecolor=blue,urlcolor=blue]{hyperref}

\renewcommand{\arraystretch}{1.2}  
\usepackage{listings}
\usepackage{color}
\definecolor{dkgreen}{rgb}{0,0.6,0}
\definecolor{gray}{rgb}{0.5,0.5,0.5}
\definecolor{mauve}{rgb}{0.58,0,0.82}
\definecolor{golden}{rgb}{0.86,0.65,0.01}
\usepackage{orcidlink}
\usepackage{hyperref}
\usepackage{amsmath}
\usepackage{amsthm}
\usepackage{amsfonts}
\usepackage{amssymb}

\usepackage{newunicodechar}
\DeclareRobustCommand{\okina}{%
  \raisebox{\dimexpr\fontcharht\font`A-\height}{%
    \scalebox{0.8}{`}%
  }%
}
\newunicodechar{ʻ}{\okina}

\newcommand{\Oumuamua}{\okina Oumuamua\xspace}
\hypersetup{pdfauthor={Name}}

\begin{document}

\title{Prompt Observations of the Interstellar Comet 3I/ATLAS from the University of Hawaii 88-inch Telescope and SNIFS: the Earliest Blue-Sensitive Spectrum and Spectrophotometric Time Series}

\title{U\lowercase{niversity} \lowercase{of} H\lowercase{awai\okina i} 88-\lowercase{inch} T\lowercase{elescope} O\lowercase{bservations} \lowercase{of} \lowercase{the} I\lowercase{nterstellar} C\lowercase{omet} 3I/ATLAS:\\ S\lowercase{pectrophotometric} B\lowercase{lue}-S\lowercase{ensitive}  S\lowercase{pectral} T\lowercase{ime} S\lowercase{eries} S\lowercase{panning} T\lowercase{wo} M\lowercase{onths} \lowercase{from} D\lowercase{iscovery}}

\author{\vspace{-1.0cm}
        W.~B.~Hoogendam$^{1*}$\orcidlink{0000-0003-3953-9532}}
\author{D.~Kuesters$^{2}$}
\author{B.~J.~Shappee$^{1}$\orcidlink{0000-0003-4631-1149}}
\author{G.~Aldering$^{3}$}
\author{J.~J.~Wray$^{4,1}$\orcidlink{0000-0001-5559-2179}}
\author{B.~Yang$^{5}$\orcidlink{0000-0002-5033-9593}}
\author{K.~J.~Meech$^{1}$\orcidlink{0000-0002-2058-5670}}

\author{M.~A.~Tucker$^{6,7,\dag}$\orcidlink{0000-0002-2471-8442}}
\author{M.~E.~Huber$^{1}$\orcidlink{0000-0003-1059-9603}}
\author{K.~Auchettl$^{8,9}$\orcidlink{0000-0002-4449-9152}}
\author{C.~R.~Angus$^{10,11}$\orcidlink{0000-0002-4269-7999}}
\author{D.~D.~Desai$^{1}$\orcidlink{0000-0002-2164-859X}}
\author{J.~T.~Hinkle$^{12,13,1\ddag}$\orcidlink{0000-0001-9668-2920}}
\author{J.~Kiyokawa$^{14}$\orcidlink{0009-0003-0398-0382}}
\author{G.~S.~H.~Paek$^{1}$\orcidlink{0000-0002-6639-6533}}
\author{S.~Romagnoli$^{8}$\orcidlink{0009-0003-8153-9576}}
\author{J.~Shi$^{8}$\orcidlink{0009-0008-3724-1824}}
\author{A.~Syncatto$^{15,1}$\orcidlink{0009-0000-6821-9285}}

\author{C.~Ashall$^{1}$\orcidlink{0000-0002-5221-7557}}
\author{M.~Dixon$^{15}$\orcidlink{0000-0003-0928-0494}}
\author{K.~Hart$^{1}$\orcidlink{0009-0003-2390-2840}}
\author{A.~M.~Hoffman$^{1}$\orcidlink{0000-0002-8732-6980}}
\author{D.~O.~Jones$^{15}$\orcidlink{0000-0002-6230-0151}}
\author{K.~Medler$^{1}$\orcidlink{0000-0001-7186-105X}}
\author{C.~Pfeffer$^{1}$\orcidlink{0000-0002-7305-8321}}

\affiliation{$^1$Institute for Astronomy, University of Hawai\okina i, Honolulu, HI 96822, USA}
\affiliation{$^2$Deutsches Elektronen-Synchrotron, Platanenallee 6, 15738 Zeuthen, Germany}
\affiliation{$^3$Lawrence Berkeley National Laboratory, 1 Cyclotron Road, MS 50B-4206, Berkeley, CA 94720, USA}
\affiliation{$^4$School of Earth and Atmospheric Sciences, Georgia Institute of Technology, 311 Ferst Drive, Atlanta, GA 30332, USA}
\affiliation{$^5$Instituto de Estudios Astrof\'isicos, Facultad de Ingenier\'ia y Ciencias, Universidad Diego Portales, Santiago, Chile}
\affiliation{$^6$Center for Cosmology \& Astroparticle Physics, The Ohio State University, Columbus, OH, USA}
\affiliation{$^7$Department of Astronomy, The Ohio State University, Columbus, OH, USA}
\affiliation{$^8$School of Physics, University of Melbourne, Parkville, VIC 3010, Australia}
\affiliation{$^9$Department of Astronomy and Astrophysics, University of California, Santa Cruz, CA 93105, USA}
\affiliation{$^{10}$Astrophysics Research Centre, School of Mathematics and Physics, Queen’s University Belfast, Belfast BT7 1NN, UK}
\affiliation{$^{11}$DARK, Niels Bohr Institute, University of Copenhagen, Jagtvej 128, DK-2200 Copenhagen {\O} Denmark}
\affiliation{$^{12}$Department of Astronomy, University of Illinois Urbana-Champaign, 1002 West Green Street, Urbana, IL 61801, USA}
\affiliation{$^{13}$NSF-Simons AI Institute for the Sky (SkAI), 172 E. Chestnut St., Chicago, IL 60611, USA}
\affiliation{$^{14}$Department of Astronomy, University of Wisconsin, Madison, WI 53706, USA}
\affiliation{$^{15}$Institute for Astronomy, University of Hawai\okina i, 640 N.~Aʻohoku Pl., Hilo, HI 96720, USA}

\altaffiliation{$^*$NSF Fellow}
\altaffiliation{$^\dag$CCAPP Fellow}
\altaffiliation{$^\ddag$NHFP Einstein Fellow}

\begin{abstract}
Interstellar objects are the ejected building blocks of other solar systems. As such, they enable the acquisition of otherwise inaccessible information about nascent extrasolar systems. The discovery of the third interstellar object, 3I/ATLAS, provides an opportunity to explore the properties of a small body from another solar system and to compare it to the small bodies in our own. To that end, we present spectrophotometric observations of 3I/ATLAS taken using the SuperNova Integral Field Spectrograph on the University of Hawai\okina i~2.2-m telescope. Our data includes the earliest $\lambda\leq3800$~\AA\ spectrum of 3I/ATLAS, obtained $\sim$12.5 hours after the discovery announcement. Later spectra confirm previously reported cometary activity, including Ni and CN emission. The data show wavelength-varying spectral slopes ($S\approx($0\%--29\%)/1000 \AA, depending on wavelength range) throughout the pre-perihelion ($r_h=4.4$--$2.5$~au) approach of 3I/ATLAS. We perform synthetic photometry on our spectra and find 3I/ATLAS shows mostly stable color evolution over the period of our observations, with $g-r$ colors ranging from $\sim$0.69--0.75~mag, $r-i$ colors ranging from $\sim$0.26--0.30~mag, and $c-o$ colors ranging from $\sim$0.50--0.55~mag. Ongoing post-perihelion observations of 3I/ATLAS will provide further insight into its potentially extreme composition.  
\end{abstract}


\section{Introduction}\label{sec:intro}
Comets and asteroids from other solar systems that pass through our own solar system on hyperbolic orbits are nominally common \citep[e.g.,][]{Engelhardt2017, Do2018}, but they are rarely observed. Once discovered, these objects offer a unique and exclusive glimpse into the small bodies of other stellar systems and the physical and chemical processes that lead to their formation \citep[e.g.,][]{Jewitt2023ARAA}.  Whereas solar system comets enable the study of primordial material from our own solar system (e.g., \citealp{Bodewits2024}), interstellar objects perform an analogous function: they enable the study of the primordial material for other solar systems \citep[e.g.,][]{Fitzsimmons2024}. Recently, \citet[][]{Denneau2025} reported the discovery of a third interstellar object by the Asteroid Terrestrial-impact Last Alert System (ATLAS; \citealp{Tonry2018a, Tonry2025}). This object, named 3I/ATLAS, is only the third interstellar object ever discovered, following 1I/\Oumuamua \citep{Meech2017} and 2I/Borisov \citep{borisov_2I_cbet}.

The first two interstellar objects showed disparate properties. Cometary activity, arising from solar radiation heating the surface-layer materials \citep[e.g.,][]{Whipple1950, Whipple1951}, differed between the previous two interstellar objects, with 1I/\Oumuamua lacking observed cometary activity \citep[e.g.,][]{Meech2017, Ye2017, Jewitt2017, ISSI_1I_review, Trilling2018} and 2I/Borisov showing strong outgassing activity \citep{Fitzsimmons:2019, Jewitt2019b, Cremonese2020, Guzik:2020, Hui2020, Kim2020, McKay2020, ye2020_borisov, yang2021} and a composition mirroring solar-system comet-like features such as $\mathrm{C}_2$ \citep{Lin2020}, [O~\textsc{I}] \citep{McKay2020}, OH \citep{Xing2020}, $\mathrm{NH}_2$ \citep{Bannister2020} species, and Ni outgassing \citep{Guzik2021, Opitom2021}. While similar to Solar System comets in many respects, 2I/Borisov had an unusually high abundance of CO \citep{Bodewits2020, Cordiner2020}. Both 1I/\Oumuamua and 2I/Borisov had similar reflectance spectra, with spectral slopes ranging from 5\% to 25\% per 1000~\AA, depending on the wavelength range and data source \citep[for a review, see][]{Jewitt2023ARAA}. 

The recently discovered 3I/ATLAS shows cometary activity like 2I/Borisov \citep{Seligman2025, Jewitt2025, Cordiner2025, Frincke25, Rahatgaonkar2025, Opitom2025, delaFuenteMarcos2025, Chandler2025, Lisse2025, Paek26, Medler26}. The initial spectrum from the University of Hawai\okina i 2.2-meter (UH~2.2m) telescope and other facilities exhibited a red-sloped reflectance spectrum without strong emission features \citep{Seligman2025, Opitom2025, Puzia2025}. As it approached perihelion, outgassing increased. Absorption by water ice grains \citep{Yang2025} and emission from CN \citep{Rahatgaonkar2025, Hoogendam25_KCWI, SalazarManzano2025, Medler26}, HCN \citep[][see also \citealt{Hinkle_JCMT}]{Coulson25}, Ni \citep{Rahatgaonkar2025, Hoogendam25_KCWI}, Fe \citep{Hutsemekers25}, $\mathrm{CO}_2$ \citep{Lisse2025, Cordiner2025}, CO \citep{Cordiner2025}, and OH \citep{Xing2025} have been reported pre-perihelion, with $\mathrm{C}_2$, $\mathrm{C}_3$, and CH reported post-perihelion \citep{Hoogendam26_KCWI}. 

Unlike previous comets, 3I/ATLAS is benefiting from intensive integral field unit (IFU) observations. IFUs offer two advantages over traditional slit spectrographs for comet studies: they enable spatially resolved observations and spectrophotometric flux calibration. Previously reported IFU data on 3I/ATLAS included a spectrum from \citet{Opitom2025}, taken shortly ($\sim$3 hours) after the discovery announcement, by the VLT-mounted Multi Unit Spectroscopic Explorer (MUSE, \citealp{Bacon10_MUSE}). Unfortunately, MUSE lacks the blue-wavelength coverage needed to observe common comet emission features, such as CN. \citet{Opitom2025} used the Wide Field Mode with a 1\arcmin~by~1\arcmin~field of view and a wavelength range of 4800--9200~\AA. Additionally, \citet{Hoogendam25_KCWI} presented a Keck Cosmic Web Imager spectrum \citep{Morrissey18} obtained nearly two months after discovery. KCWI can observe blue wavelengths, and these data reveal that Ni emission is more centrally concentrated than CN emission and broadband coma light, suggesting Ni is likely released directly from the nucleus. This contrasts with longer-lived second-generation species such as CN, which have much longer lifetimes against photodissociation.

In this manuscript, we present a spectral time series of integral-field unit (IFU) observations of 3I/ATLAS from the SuperNova Integral Field Spectrograph (SNIFS; \citealp{Lantz2004}) on the University of Hawai\okina i (UH) 2.2m telescope, taken as part of the Spectroscopic Classification of Astronomical Transients (SCAT; \citealp{Tucker2022}) survey.

\section{Data}\label{sec:data}

Nine spectra were obtained between UT 2025 July 3 and UT 2025 September 2 using the SNIFS instrument on the UH~2.2-meter telescope at Maunakea by the SCAT team. A log of the data is presented in Table~\ref{tab:spec}. The SCAT collaboration uses the UH~2.2-meter and the Australian National University~2.3-meter telescopes to classify and spectroscopically follow transient phenomena such as supernovae \citep[e.g.,][]{Tucker24_ufx, Hoogendam25_epr, Hoogendam25_pxl}, tidal disruption events \citep[e.g.,][]{Hinkle21_dj, Hinkle23_mlx, Hoogendam24_TDE, Hinkle24_ci, Hinkle25_ENT, Pandey25}, active galactic nuclei \citep[e.g.,][]{Neustadt23}, X-ray binaries \citep[e.g.,][]{Tucker18}, and dipper stars \citep[e.g.,][]{Fores-Toribio25}. 

SNIFS contains two channels split by a dichroic mirror. The blue channel covers the 3400--5100~\AA\ range, and the red channel covers 5100--10000~\AA. The spectral resolutions are 5 \r{A} and 7 \r{A}, respectively. The data were reduced using the Nearby Supernova Factory \citep{Aldering02} reduction pipeline \citep{Bacon01, Aldering06, Scalzo10}, and flux calibrated following the steps detailed in \citet{Buton13}. Fluxes were extracted using the Gaussian+Moffat point-spread-function model\footnote{It is not required to be a point source, but the proportion and Moffat $\beta$ are fixed to work best for point sources.} with a flat background, as described in \citet{Rubin2022}. The coma of 3I/ATLAS is the dominant flux source \citep[e.g.,][]{Jewitt2025}, even pre-discovery \citep[e.g.,][]{Ye2025}. Thus, some of the coma may end up in the sky spectrum computed by the SNF pipeline. At early epochs, this effect is likely negligible, as confirmed by our inspection of the cubes. However, this is not the case for later epochs.

For the last three epochs, 3I/ATLAS fills the extent of the cube. Therefore, the SNF pipeline inadvertently includes comet flux in the sky spectrum, resulting in cometary self-subtraction. Indeed, inspection of the spectral cubes confirms the presence of CN emission and a small, flat, nonzero continuum contribution in the sky spectrum. The limited continuum contamination at these later epochs, when 3I/ATLAS fills the SNIFS field of view, suggests that the previously noted potential continuum contamination in early epochs is indeed likely negligible. To remedy contamination in the CN line flux, we perform aperture extraction on the cubes at 2\arcsec and 3\arcsec and use supernova observations from the same night to create uncontaminated sky spectra for the final three epochs of spectroscopy. We use these decontaminated spectra in Section \ref{sec:activity} to compute production rates for 3I/ATLAS.

The solar analogue HD~165290 was observed at all epochs except 2025-07-03 and 2025-07-19. For those epochs, we use observations of HD~165290 from the preceding or following night, respectively. For the final epoch on 2025-09-02, HD~165290 was unobservable due to a small lunar separation of 4 degrees, so the solar analogue HD~142801 was observed instead. We compared the spectra of HD~142801 and HD~165290 and find a negligible difference after correcting for the magnitude difference between the two stars. Each final calibrated reflectance spectrum was obtained by dividing the target spectrum by the solar analogue spectrum and normalizing at 5500 \AA. Because we use solar analogues with similar spectral shapes\footnote{all but one epoch use the same solar analogue} for all our observations, differences in spectral slope evolution will not be introduced by the solar analogue.

\begin{deluxetable}{ccccccc}
\tablenum{1}
\tablecaption{Log of spectroscopic observations. $r_h$ is the heliocentric distance, $\Delta$ is the geocentric distance, $\alpha$ is the phase angle, and TA is the true anomaly, or position along the orbit.}\label{tab:spec}
\tablewidth{\linewidth}
\tablehead{ \colhead{UT Date} & \colhead{MJD} &  \colhead{$r_h$} & \colhead{$\Delta$} & \colhead{$\alpha$} & \colhead{TA} \\
            \colhead{} & \colhead{} &  \colhead{au} & \colhead{au} & \colhead{Degrees} &\colhead{Degrees}}
\startdata
2025-07-03  &    60859.3 &   4.43   &   3.44  &  2.6 & -78.9\\
2025-07-04  &    60860.4 &   4.40   &   3.40  &  3.0 & -78.7\\
2025-07-12  &    60868.4 &   4.14   &   3.19  &  5.9 & -77.4\\
2025-07-14  &    60870.4 &   4.07   &   3.14  &  6.6 & -77.0\\
2025-07-18  &    60874.3 &   3.94   &   3.06  &  8.2 & -76.3\\
2025-07-19  &    60875.3 &   3.91   &   3.03  &  8.6 & -66.1\\
2025-08-18  &    60905.3 &   2.94   &   2.64  & 20.0 & -68.0\\
2025-08-23  &    60910.2 &   2.78   &   2.61  & 21.3 & -66.2\\
2025-09-02  &    60920.6 &   2.47   &   2.57  & 23.0 & -61.6
\enddata
\end{deluxetable}

\begin{figure*}
\includegraphics[width=\textwidth]{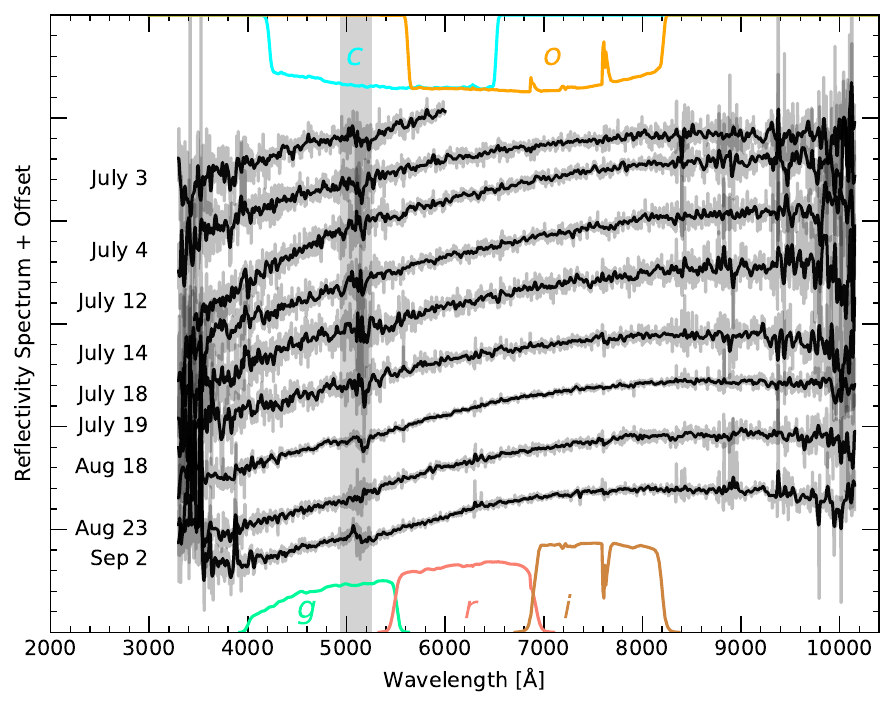}
\caption{The spectrophotometric time series data of 3I/ATLAS from the UH~2.2-meter telescope taken with the SNIFS spectrograph. Grey curves are the original data, and black curves represent the Gaussian-smoothed data. The SNIFS dichroic boundary between $\sim$5000-5200~\AA\ is shaded grey. The Nearby Supernova Factory pipeline improves the dichroic region; however, some spectra still exhibit low signal-to-noise ratios there. The Pan-STARRS (bottom) and ATLAS (top) filter transmission functions are also shown. The red channel of the first spectrum is contaminated by a stellar streak, leading to a slope that differs from those of the other spectra and is therefore excluded. The blue channel is unaffected by the stellar streak because the contaminating star is extincted by the Milky Way.}
\label{fig:3I_SNIFS}
\end{figure*} 

\begin{deluxetable}{cccccc}
\tablenum{2}
\tablecaption{Log of synthetic photometry. Magnitudes are reported in the AB system. See \citet{Tonry2018a} for $co$ bandpass information and \citet{Chambers2016} for $griz$ bandpass information. }\label{tab:syn_phot_table}
\tablewidth{\linewidth}
\tablehead{ \colhead{UT Date} & \colhead{MJD} & \colhead{$r_h$} & \colhead{Filter} &  \colhead{Magnitude} & \colhead{Magnitude Error} }
\startdata
2025-07-04 & 60860.4 & 4.40 & $g$ & 18.72 & 0.01 \\
2025-07-04 & 60860.4 & 4.40 & $c$ & 18.41 & 0.01 \\
2025-07-04 & 60860.4 & 4.40 & $r$ & 18.03 & 0.01 \\
2025-07-04 & 60860.4 & 4.40 & $o$ & 17.91 & 0.01 \\
2025-07-04 & 60860.4 & 4.40 & $i$ & 17.77 & 0.01 \\
2025-07-04 & 60860.4 & 4.40 & $z$ & 17.72 & 0.01 \\
2025-07-12 & 60868.4 & 4.14 & $g$ & 18.54 & 0.01 \\
2025-07-12 & 60868.4 & 4.14 & $c$ & 18.20 & 0.01 \\
2025-07-12 & 60868.4 & 4.14 & $r$ & 17.80 & 0.01 \\
2025-07-12 & 60868.4 & 4.14 & $o$ & 17.66 & 0.01 \\
2025-07-12 & 60868.4 & 4.14 & $i$ & 17.51 & 0.01 \\
2025-07-12 & 60868.4 & 4.14 & $z$ & 17.42 & 0.01 \\
2025-07-14 & 60870.4 & 4.07 & $g$ & 18.43 & 0.01 \\
2025-07-14 & 60870.4 & 4.07 & $c$ & 18.09 & 0.01 \\
2025-07-14 & 60870.4 & 4.07 & $r$ & 17.69 & 0.01 \\
2025-07-14 & 60870.4 & 4.07 & $o$ & 17.54 & 0.01 \\
2025-07-14 & 60870.4 & 4.07 & $i$ & 17.39 & 0.01 \\
2025-07-14 & 60870.4 & 4.07 & $z$ & 17.29 & 0.01 \\
2025-07-18 & 60874.4 & 3.94 & $g$ & 18.20 & 0.01 \\
2025-07-18 & 60874.4 & 3.94 & $c$ & 17.89 & 0.01 \\
2025-07-18 & 60874.4 & 3.94 & $r$ & 17.52 & 0.01 \\
2025-07-18 & 60874.4 & 3.94 & $o$ & 17.37 & 0.01 \\
2025-07-18 & 60874.4 & 3.94 & $i$ & 17.23 & 0.01 \\
2025-07-18 & 60874.4 & 3.94 & $z$ & 17.14 & 0.01 \\
2025-07-19 & 60875.4 & 3.91 & $g$ & 18.14 & 0.01 \\
2025-07-19 & 60875.4 & 3.91 & $c$ & 17.83 & 0.01 \\
2025-07-19 & 60875.4 & 3.91 & $r$ & 17.46 & 0.01 \\
2025-07-19 & 60875.4 & 3.91 & $o$ & 17.33 & 0.01 \\
2025-07-19 & 60875.4 & 3.91 & $i$ & 17.21 & 0.01 \\
2025-07-19 & 60875.4 & 3.91 & $z$ & 17.17 & 0.01 \\
2025-08-18 & 60905.3 & 2.94 & $g$ & 17.41 & 0.01 \\
2025-08-18 & 60905.3 & 2.94 & $c$ & 17.09 & 0.01 \\
2025-08-18 & 60905.3 & 2.94 & $r$ & 16.71 & 0.01 \\
2025-08-18 & 60905.3 & 2.94 & $o$ & 16.57 & 0.01 \\
2025-08-18 & 60905.3 & 2.94 & $i$ & 16.42 & 0.01 \\
2025-08-18 & 60905.3 & 2.94 & $z$ & 16.35 & 0.01 \\
2025-08-23 & 60910.2 & 2.78 & $g$ & 17.28 & 0.01 \\
2025-08-23 & 60910.2 & 2.78 & $c$ & 16.93 & 0.01 \\
2025-08-23 & 60910.2 & 2.78 & $r$ & 16.52 & 0.01 \\
2025-08-23 & 60910.2 & 2.78 & $o$ & 16.38 & 0.01 \\
2025-08-23 & 60910.2 & 2.78 & $i$ & 16.23 & 0.01 \\
2025-08-23 & 60910.2 & 2.78 & $z$ & 16.17 & 0.01 \\
2025-09-02 & 60920.6 & 2.47 & $g$ & 16.86 & 0.01 \\
2025-09-02 & 60920.6 & 2.47 & $c$ & 16.54 & 0.01 \\
2025-09-02 & 60920.6 & 2.47 & $r$ & 16.16 & 0.01 \\
2025-09-02 & 60920.6 & 2.47 & $o$ & 16.02 & 0.01 \\
2025-09-02 & 60920.6 & 2.47 & $i$ & 15.88 & 0.01 \\
2025-09-02 & 60920.6 & 2.47 & $z$ & 15.82 & 0.01 \\
\enddata
\end{deluxetable}

\section{Analysis}\label{sec:analysis}

Figure \ref{fig:3I_SNIFS} shows the spectrophotometric time series data presented in this work. The July 3rd spectrum shows a stellar streak that overlaps with the comet, resulting in a different spectral slope in the redder bands. The contaminating stellar source is extincted by Milky Way dust, reducing the contaminating flux at bluer wavelengths. This makes wavelengths below $\sim$6000~\AA\ mostly contamination-free, and our visual inspection of the cubes confirms that 3I/ATLAS is the dominant source of flux at these wavelengths. However, at redder wavelengths the contamination flux is non-negligible, so we mask wavelengths redder than $\sim$6000~\AA\ and exclude this spectrum from our spectral slope analysis below.

\subsection{Synthetic Photometry}
\begin{figure}
\includegraphics[width=\linewidth]{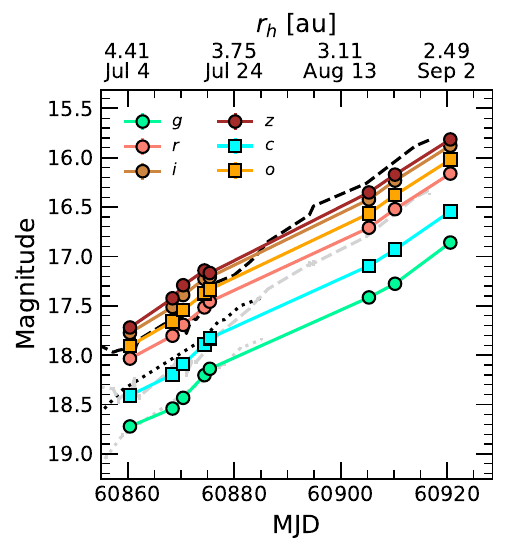}
\caption{Synthetic photometry in the Pan-STARRS $griz$ and ATLAS $co$ passbands. All magnitudes are reported in the AB system. Reported ``m10'' ATLAS magnitudes from \citet{Tonry2025} are shown as black lines, and the ``m6'' ATLAS magnitudes are shown as grey lines. The cyan filter is denoted using dotted lines, and the orange filter is denoted using dashed lines. }
\label{fig:synth_phot}
\end{figure}

The accurate spectral flux calibration of our SNIFS spectra enables reliable synthetic photometry. The SNF pipeline source extraction prevents exact knowledge of the aperture, but visual inspection of the cubes suggests a majority of the flux is contained within a 2--3\arcsec radius, which would be consistent with the extraction radius used in other IFU studies \citep[e.g.,][]{Hoogendam25_KCWI}. Table~\ref{tab:syn_phot_table} contains a log of our synthetic photometry. The synthetic Pan-STARRS $gri$ \citep{Chambers2016} and ATLAS $co$ \citep{Tonry2018a} photometry from photometric nights is plotted in Figure \ref{fig:synth_phot}, along with previously reported ATLAS photometry from \citet[][]{Tonry2025}. Differences in magnitude are expected due to the different apertures used, which will contain different amounts of the cometary coma. Shortly after discovery, the $c$- and $o$-band data agree with our synthetic magnitudes, likely due to the cometary activity being entirely encapsulated in both the ATLAS aperture and SNIFS field-of-view. As cometary activity increases, the larger ATLAS apertures capture more cometary flux and thus appear brighter than our values.

3I/ATLAS shows a stable color evolution across most of our observations, with observed $g-r$ colors ranging from 0.69 to 0.75~mag, $r-i$ colors from 0.26 to 0.30~mag, and $c-o$ colors from 0.50 to 0.55~mag. The July 12 and 14 spectra differ in color, but there may be a poorly understood data issue with the July 12 data (see Section \ref{sec:slopes} for further discussion). The measured colors from different groups exhibit systematic variations, likely due to differences in filter bandpasses and aperture sizes, yet each reported color is consistent with a moderately red, dust-dominated coma.  \citet{Puzia2025} measured SDSS $g-r = 0.86\pm0.05$~mag and Pan-STARRS $g-r = 0.73\pm0.05$~mag from July 4 flux-calibrated spectra, and  \citet{Kareta2025} report $g'-i' = 0.98\pm0.03$~mag from IRTF observations using \okina Opihi \citep{Lee2022_Opihi}, whereas our PSF-extracted IFU data yield slightly bluer values for the Pan-STARRS filters ($g-r\approx0.69$--0.75~mag).  \citet{Beniyama25} report slightly bluer colors ($g-r\approx0.60$~mag, $r-i\approx0.21$~mag) on July 15, and   \citet{Bolin2025} reported $B-V = 0.98\pm0.23$~mag, $V-R = 0.71\pm0.09$~mag, $g-r = 0.84\pm 0.05$~mag, $r-i = 0.16\pm0.03$~mag.

\subsection{Activity}\label{sec:activity}

\begin{deluxetable*}{ccccccc}
\tablenum{3}
\tablecaption{Log of CN production rates. Table~\ref{tab:spec} contains orbital properties like heliocentric and geocentric distances. The rightmost column reports the preferred production rates using spectra corrected for potential self-subtraction (black spectra in Figure \ref{fig:activity}). }\label{tab:prod_rate}
\tablewidth{\linewidth}
\tablehead{\colhead{UT Date} & \colhead{MJD} & \colhead{Flux} & \colhead{3\arcsec\ Rate}               & \colhead{2\arcsec\ Rate}               & \colhead{2\arcsec\ Rate Corrected} \\
           \colhead{}        & \colhead{}    & \colhead{erg\,cm$^{-2}$\,s$^{-1}$}    & \colhead{$10^{24}$ molecules~s$^{-1}$} & \colhead{$10^{24}$ molecules~s$^{-1}$} & \colhead{$10^{24}$ molecules~s$^{-1}$}}
\startdata
2025-07-12  &  60868.4  & $\dots$ & $< 0.98$       &  $\cdots$      &  $\cdots$ \\
2025-08-18  &  60905.3  & $(4.44\pm0.68)\times10^{-15}$  &  $0.8 \pm 0.3$  &  $1.8 \pm 0.7$ &  $8.2\pm1.2$ \\
2025-08-23  &  60910.2  & $(9.41\pm1.84)\times10^{-15}$ &  $1.2 \pm 0.3$  &  $2.5 \pm 0.6$ &  $14\pm3$ \\
2025-09-02  &  60920.6  & $(3.25\pm0.56)\times10^{-14}$ &  $3.1 \pm 0.5$  &  $6.5 \pm 1.1$ &  $29\pm5$
\enddata
\end{deluxetable*}

\begin{figure*}
\includegraphics[width=\textwidth]{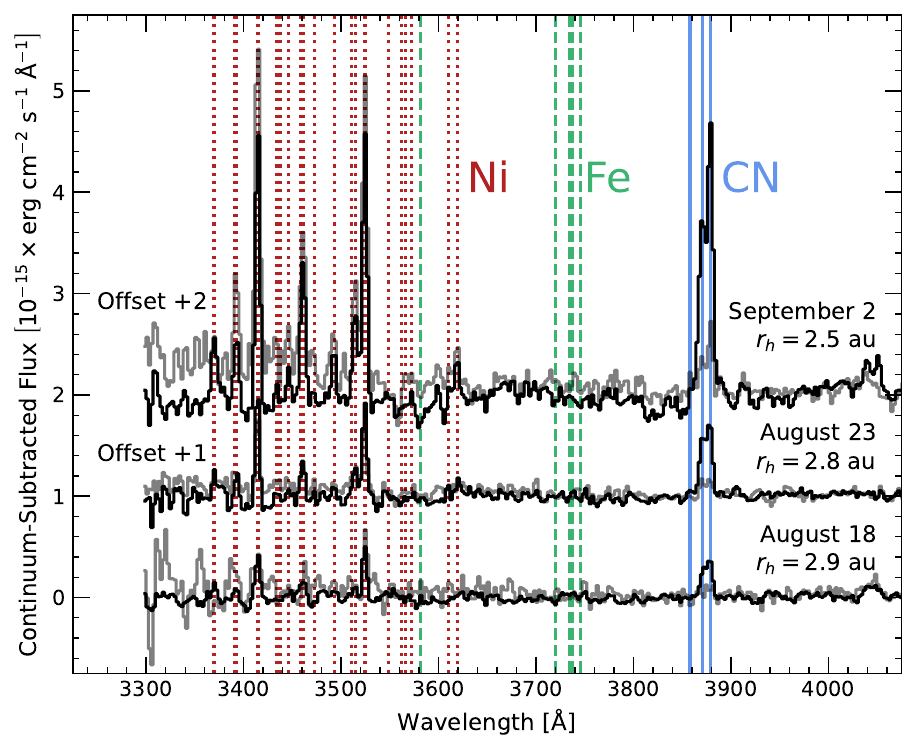}
\caption{The continuum-subtracted spectra showing emission lines from Ni (dotted red), Fe (dashed green), and CN (solid blue). The grey spectra use the default sky subtraction routine in the SNF pipeline, and the black spectra use a sky subtraction based on non-3I/ATLAS observations from the same night (see text for details).}
\label{fig:activity}
\end{figure*}

The final three epochs of SNIFS spectra contain UV emission lines arising from cometary activity (see Figure \ref{fig:activity}). The flux of UV emission lines can be spectrally isolated, avoiding potential oversubtraction of sky flux. This is possible because the reflected comet light, the zodiacal light that dominates the terrestrial sky background at dark sites like Mauna Kea, and moonlight (present on some nights), all have continuum spectral shapes dominated by scattered sunlight. Some night sky emission lines are present, but are weak compared to the continuum and are at known wavelengths that do not significantly contaminate cometary CN emission \citep[$\sim 3$\% ][]{Hanuschik2003}. This means that the spectra from our solar analogues can be warped, scaled, and subtracted to isolate the CN emission, similar to \citet{Rahatgaonkar2025}, \citet{Hoogendam25_KCWI}, and \citet{Hoogendam26_KCWI}. Accordingly, we model the continuum plus sky contribution as a function of wavelength as:

\begin{equation}
    F_{\mathrm{cont}}(\lambda) = R(\lambda)\times F_{\odot\,\mathrm{analogue}}\left[\lambda\left(1+\frac{v}{c}\right)+\delta\lambda\right].
\end{equation}

\noindent The function $R(\lambda)$ is a fifth-order polynomial fit over the full wavelength range of the SNIFS blue channel. $F_{\odot\,\mathrm{analogue}}$ is the shifted solar analogue spectrum. The shift parameters, $v$ and $\delta\lambda$, account for the relative motion between Earth and the comet and between Earth and the solar analogue star.  The high signal-to-noise ratio of the solar analogue data results in a negligible contribution to the overall error budget.

The resulting background-subtracted spectra spatially integrated over the field of view are shown in Figure \ref{fig:activity}. It is apparent that the emission lines are well isolated, with very little evidence of background continuum or night-sky residuals. We have also verified that the resulting spectra from nights without cometary emission are flat and featureless. Now that we have a data cube containing only the residual cometary emission lines, we measure the aperture flux in 2\arcsec and 3\arcsec apertures centered on the comet as input to production rate calculations. As portions of the 3\arcsec aperture fell outside the field, circular symmetry was assumed to estimate the missing flux.

We observe CN and \ion{Ni}{1} features that have been previously reported \citep[e.g.,][]{Rahatgaonkar2025, Hoogendam25_KCWI}. Additionally, \citet{Hutsemekers25} report \ion{Fe}{1} features emerging after the CN and Ni emission on August 28 and stacked September 3+4. Our September 2nd spectrum may show emission consistent with the reported \ion{Fe}{1} features, but the signal-to-noise ratio of the lines is insufficient to confirm a clear detection. 

A simple \citep{Haser:1957} model was used to convert the measured CN line flux into a gas production rate. The scale lengths were taken from \citet{A'Hearn:1995} and the number of photons emitted per molecule per second (the so-called g-factor) is taken from \citet{Schleicher:2010}, accounting for the Swings effect and scaled by $r_h^{-2}$. We assumed that the gas escapes isotropically from the nucleus at a constant velocity, and adopted a mean expansion speed of $0.8 \times r_h^{-0.5}$~km~s$^{-1}$, following \citet{Biver:1999}, where $r_h$ is the heliocentric distance in au.

Because the spectral extraction method for the SNIFS data cubes relies on a PSF model over the entire cube, the exact aperture (or the effective aperture weighted by the PSF that results in an ``equivalent noise area'' \citealp{King83}) over which the spectrum was extracted is non-trivial. Because the angular extent of 3I/ATLAS may vary, such an aperture is not constant and is difficult to use for literature comparisons. Instead, we report inferred production rates measured from the PSF-extracted spectra assuming two different aperture radii: 2\arcsec\ and 3\arcsec. While 2\arcsec\ or 3\arcsec\ apertures are not constant with respect to the ``equivalent noise area'', they at least facilitate a more straightforward literature comparison. On the other hand, the cubes using the manual sky subtraction are extracted with a known aperture of 2\arcsec. In this case, we report just the 2\arcsec values. We recommend that meta-analyses and future works use these values instead of the other values, both because of the known aperture and because of the improved sky subtraction. Our CN production rates are logged in Table~\ref{tab:prod_rate}. 

We also report an upper limit of $\mathrm{Q(CN)}< 9.8 \times10^{23}$~molecules~s$^{-1}$ from our highest SNR spectrum, taken on July 12. Our first detection is on August 18 with a production rate of $\left(8.4\pm3.2\right)\times10^{23}$~molecules~s$^{-1}$ assuming a 3\arcsec\ effective aperture radius. Finally, we compute $Af\rho$ values assuming a 3\arcsec\ aperture using the spectral range between 6000–6200~\AA, a region largely free of gas contamination and close to the $R$ band. Our July $Af\rho$ values are consistent with \citet{Santana-Ros2025}, but our August $Af\rho$ values are much higher than the $r$-band values reported by \citet{SalazarManzano2025}, likely reflecting differences in aperture size and wavelength range. It is possible that the dust coma of 3I is not in a steady state and that a smaller aperture yields a higher $Af\rho$ value.

\begin{deluxetable}{ccccccc}
\tablenum{1}
\tablecaption{Tabulated $Af\rho$ values for 3I/ATLAS. $r_h$ is the heliocentric distance, and $\Delta$ is the geocentric distance. Af$\rho$ is the dust production, and $\sigma$Af$\rho$ is its uncertainty.}\label{tab:afrho}
\tablewidth{\linewidth}
\tablehead{ \colhead{UT Date} & \colhead{MJD} &  \colhead{$r_h$} & \colhead{$\Delta$} & \colhead{$Af\rho$} & \colhead{$\sigma$Af$\rho$} \\
            \colhead{} & \colhead{} &  \colhead{au} & \colhead{au} & \colhead{cm} &\colhead{cm}}
\startdata
2025-07-03  &    60859.3 &   4.43   &   3.44  & 329 & 16\\
2025-07-04  &    60860.4 &   4.40   &   3.40  & 306 & 14\\
2025-07-12  &    60868.4 &   4.14   &   3.19  & 329 & 14\\
2025-07-14  &    60870.4 &   4.07   &   3.14  & 312 & 20\\
2025-07-18  &    60874.3 &   3.94   &   3.06  & 414 & 22\\
2025-07-19  &    60875.3 &   3.91   &   3.03  & 437 & 16\\
2025-08-18  &    60905.3 &   2.94   &   2.64  & 613 &  9\\
2025-08-23  &    60910.2 &   2.78   &   2.61  & 669 & 19\\
2025-09-02  &    60920.6 &   2.47   &   2.57  & 773 & 15
\enddata
\end{deluxetable}

\subsection{Spectral Slopes}\label{sec:slopes}
\begin{figure}
\includegraphics[width=\linewidth]{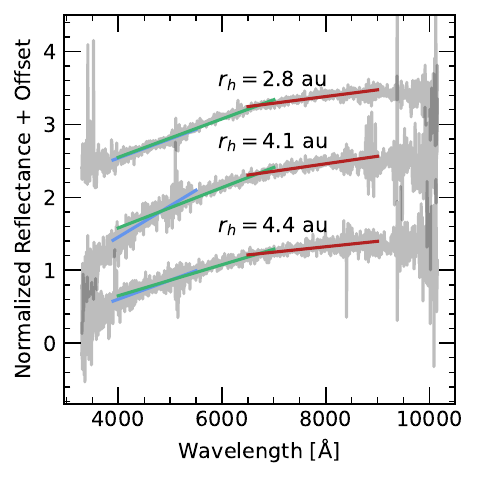}
\caption{The spectral slope evolution fits for several epochs of 3I/ATLAS SNIFS spectra. The different colors represent different fitted wavelength ranges.}
\label{fig:slope_fits}
\end{figure}

\begin{figure}
\includegraphics[width=\linewidth]{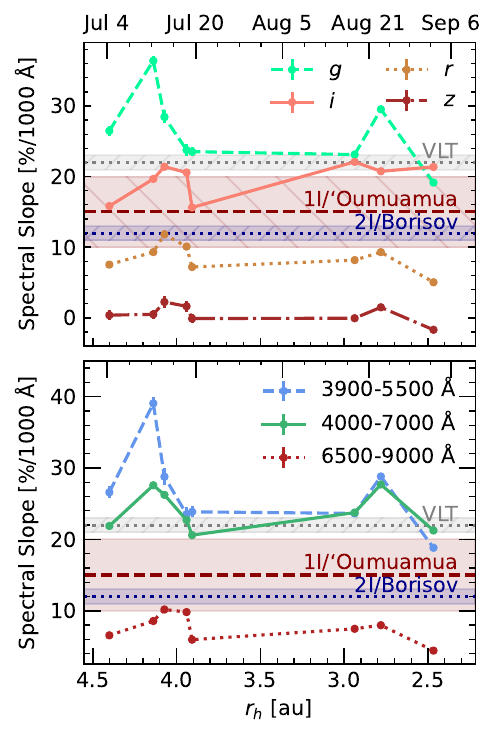}
\caption{The spectral slope evolution for 3I/ATLAS from our SNIFS spectra. The slope is described in the text. We compute the slopes using a linear fit to the reflectance spectra, with values normalized at 5500~\AA. The first spectrum is excluded because it lacks uncontaminated red-wavelength coverage. The crosshatched regions centered on the dashed red and dotted blue lines denote the aggregate slope values for 1I/\Oumuamua and 2I/Borisov, respectively, from \citet{Jewitt2023ARAA}. The grey-shaded region shows the spectral slope reported by \citet{Rahatgaonkar2025} for their spectroscopic time-series data from $r_h\simeq4.4$ to 2.9~au. Only statistical uncertainties are plotted. See Section \ref{sec:slopes_discussion} for discussion about the steep red slopes between July 12-14.}
\label{fig:slope_evolution}
\end{figure}

For our spectral slope calculations, we divide our spectra by the solar analog spectrum to obtain reflectance spectra, which are normalized to 1 at 5500~\AA. We measure the spectral slope for each epoch of \mbox{3I/ATLAS} using a linear fit to the normalized reflectance spectra over various wavelength ranges, masking wavelengths near the SNIFS dichroic crossover. To ensure robust error estimates, we perform 100 bootstrap iterations, randomly sampling 90\% of the spectral points in each. Figure \ref{fig:slope_fits} shows the spectral slope fits, and Figure \ref{fig:slope_evolution} shows the spectral slope evolution. 

3I/ATLAS exhibits a red reflectance color throughout its spectral evolution, likely due to its dusty coma \citep{Seligman2025, Opitom2025, Yang2025, Kareta2025, Santana-Ros2025}. From 4000~\AA\ to 7000~\AA, the slope of 3I/ATLAS ranges from 20\%/1000~\AA\ to 29\%/1000~\AA. Previously reported spectral slope values for 3I/ATLAS are consistent with our results and are compiled in Table~\ref{tab:slopes}.

\begin{deluxetable}{ccc}
\tablenum{4}
\tablecaption{Log of literature spectral slope measurements for 3I/ATLAS.}\label{tab:slopes}
\tablewidth{\linewidth}
\tablehead{\colhead{$r_h$}  & \colhead{Wavelength}  & \colhead{Slope} \\
           \colhead{(au)}   & \colhead{(\AA)}       & \colhead{(\%/1000\AA)}
}
\startdata
4.40 & 3900--5500 & $26.6 \pm 0.8$\tablenotemark{a} \\
4.14 & 3900--5500 & $39.1 \pm 0.8$\tablenotemark{a} \\
4.07 & 3900--5500 & $28.8 \pm 1.1$\tablenotemark{a} \\
3.94 & 3900--5500 & $23.5 \pm 1.1$\tablenotemark{a} \\
3.91 & 3900--5500 & $23.9 \pm 0.7$\tablenotemark{a} \\
2.94 & 3900--5500 & $23.6 \pm 0.3$\tablenotemark{a} \\
2.78 & 3900--5500 & $28.8 \pm 0.5$\tablenotemark{a} \\
2.47 & 3900--5500 & $18.8 \pm 0.4$\tablenotemark{a} \\
4.40 & 4000--5500 & $27.4 \pm 1.0$\tablenotemark{c} \\
2.80 & 3900--6000 & $26.1 \pm 0.5$\tablenotemark{g} \\
\hline & \\[-1.5ex]
3.80 & 3800--9200 & $18.3 \pm 0.9$\tablenotemark{f} \\
\hline & \\[-1.5ex]
4.40 & 4000--7000 & $17.1 \pm 0.2$\tablenotemark{b} \\
4.40 & 4000--7000 & $21.9 \pm 0.2$\tablenotemark{a} \\
4.14 & 4000--7000 & $27.6 \pm 0.3$\tablenotemark{a} \\
4.07 & 4000--7000 & $26.2 \pm 0.3$\tablenotemark{a} \\
3.94 & 4000--7000 & $22.7 \pm 0.3$\tablenotemark{a} \\
3.91 & 4000--7000 & $20.6 \pm 0.2$\tablenotemark{a} \\
2.94 & 4000--7000 & $23.8 \pm 0.1$\tablenotemark{a} \\
2.78 & 4000--7000 & $27.7 \pm 0.1$\tablenotemark{a} \\
2.47 & 4000--7000 & $21.2 \pm 0.1$\tablenotemark{a} \\
\hline & \\[-1.5ex]
4.47 & 5000--7000 & $18.0 \pm 3.0$\tablenotemark{e} \\
4.40 & 5500--7000 & $26.7 \pm 0.7$\tablenotemark{c; G2V} \\
4.40 & 5500--7000 & $16.4 \pm 0.4$\tablenotemark{c; SOLSPEC} \\
4.40 & 5000--8000 & $11.4 \pm 0.2$\tablenotemark{d} \\
3.80 & 5000--7000 & $18.5 \pm 0.5$\tablenotemark{f} \\
2.80 & 5500--7000 & $14.0 \pm 1.9$\tablenotemark{g} \\
\hline & \\[-1.5ex]
4.47 & 7000--9000 & $17.0 \pm 4.0$\tablenotemark{e} \\
4.40 & 6500--9000 & $6.5 \pm 0.2$\tablenotemark{a} \\
4.40 & 7000--9000 & $5.2 \pm 0.2$\tablenotemark{d} \\
4.14 & 6500--9000 & $8.6 \pm 0.2$\tablenotemark{a} \\
4.07 & 6500--9000 & $10.2 \pm 0.3$\tablenotemark{a} \\
4.07 & 7000--9000 & $6.7 \pm 1.4$\tablenotemark{d} \\
3.94 & 6500--9000 & $9.8 \pm 0.3$\tablenotemark{a} \\
3.91 & 6500--9000 & $5.9 \pm 0.2$\tablenotemark{a} \\
3.80 & 7000--9000 & $11.6 \pm 0.4$\tablenotemark{f} \\
2.94 & 6500--9000 & $7.5 \pm 0.1$\tablenotemark{a} \\
2.80 & 7000--9000 & $5.2 \pm 0.2$\tablenotemark{g} \\
2.78 & 6500--9000 & $8.0 \pm 0.2$\tablenotemark{a} \\
2.47 & 6500--9000 & $4.4 \pm 0.1$\tablenotemark{a}
\enddata
\tablenotetext{a}{This work - UH 2.2m/SNIFS}
\tablenotetext{b}{\citet{Seligman2025} - SNIFS SCAT pipeline reduction}
\tablenotetext{c}{\citet{Puzia2025} - SOAR/Goodman HTS}
\tablenotetext{d}{\citet{Yang2025} - Gemini South/GMOS}
\tablenotetext{e}{\citet{Opitom2025} - VLT/MUSE}
\tablenotetext{f}{\citet{delaFuenteMarcos2025} - GTC/OSIRIS}
\tablenotetext{g}{\citet{SalazarManzano2025} - MDM 2.4m}
\tablecomments{Spectral slopes are normalized reflectivity gradients typically measured by linear fits to reflectance spectra. Different solar reference spectra and wavelength ranges can lead to systematic differences between measurements. Heliocentric distances ($r_h$) are approximate values at the time of observation.}
\end{deluxetable}

\section{Discussion of Spectral Slope Results}\label{sec:slopes_discussion}

We begin our discussion with remarks on the striking, sudden red slope ``spike'' from July 12 to 14 ($r_h\approx4.1$~au). The blue slope on July 12 may likely be an erroneous slope introduced during data reduction. However, we are unable to find the exact cause. The computed instrumental response and atmospheric extinction are similar to those on other nights. While 3I/ATLAS was in a crowded field at the time of observation, we rule out stellar streak contamination for two reasons. First, the July 12th spectrum (shown in Figure \ref{fig:slope_fits}) is a median of extracted spectra from the 3 uncontaminated exposures of 3I/ATLAS (a fourth exposure was removed due to a visible streak), so any influence from stellar streak contamination should be removed during the coadd. Second, we inspect the white-light images of the cubes and confirm the absence of contaminating streaks. An additional check is to compute the spectral slopes using the secondary solar analog stars. Unfortunately, the spectra of the secondary analog for July 12, HD~157842, and HD~165290 are significantly different. Our spectrum of HD~157842 from July 12 differs from those of HD~165290 from July 12 and 14, and of the secondary solar analogue HD~154805 from July 14. The spectral slopes are consistent using any of these three stars, while the spectral slope using HD~157842 is a mere 17\%/1000~\AA, far less red than any other measurement we make.

On the other hand, the July 14 spectrum shows a similar slope regardless of which solar analogue star from that night was used. The cube does not show a stellar streak. While the slope remains redder in this epoch, it is likely more reliable than the July 12 measurement, and its variation likely arises from intrinsic systematic scatter in spectral slope measurements. 

Unfortunately, there is a dearth of literature measurements during this time phase to compare our spectral slopes. \citet{Rahatgaonkar2025} report spectral slopes from 3900~\AA\ to 5550~\AA\ of $22\pm1$~\%/1000~\AA\ for all epochs, which includes an epoch on July 17, but not during the reddest slope epochs we see. As-of-yet unpublished VLT/MUSE observations on July 16 are also consistent with our 6500~\AA\ to 9000~\AA\ slope (C. Opitom, private communication). Finally, \citet{Jewitt25_preperihelion} obtained photometry from NOT/ALFOSC on July 11, but they do not report spectral slope measurements. As additional observations in this time range are reported, particularly those covering bluer wavelengths, a better understanding of this behavior may emerge. 

If this slope trend is real, the rapid change and return may arise from the sublimation of large dust grains, effectively reddening the spectra. Micron-sized icy grains were detected in the spectrum of 3I on July 14 \citep{Yang2025}.  Grains this size near 4 au can live for many days, depending on the albedo \citep{protopapa2018}. If these ices were embedded in larger refractory grains, once the ices sublimated, only the original red refractory material would be left behind, causing the spectra to return to similar original slopes.

A second, more certain result, is that the redder slopes are flatter than the bluer slopes. This is further evidence to confirm previous speculation that the spectral slopes of 3I/ATLAS may ``turn over'' at redder wavelengths \citep[e.g.,][]{Yang2025}. Indeed, between 8000~\AA\ and 10\,000~\AA\ the slopes by eye appear nearly flat, and the 6500~\AA\ to 9000~\AA\ slopes plotted in Figure \ref{fig:slope_evolution} are far less red and less steep than the bluer wavelength ranges. While similar to the \citet{DeMeo2009} D-type spectral class, those objects do not turn over. The evidence is mounting that 3I/ATLAS turns over in the red wavelengths, as shown by this work and others (e.g., \citealp{Yang2025}), which is within the realm of typical cometary behavior \citep[e.g.,][]{Jewitt86}.

The previous two interstellar objects also had red slopes. The spectral slope for 1I/\Oumuamua ranged from 7~to~23\%/1000~\AA\ \citep[e.g.,][]{Jewitt2017, Meech2017, Ye2017, Fitzsimmons2018}, with an average value of $(15\pm5)$ \%/1000~\AA\ \citep{Jewitt2023ARAA}. Unlike 2I/Borisov and 3I/ATLAS, the spectral slope of 1I/\Oumuamua reflects the bare nucleus instead of a dusty coma. Of the slope values for 1I/\Oumuamua, the most similar method is from \citet{Meech2017}, who removed the rotation light curve and reported a spectral slope of $\sim$23\%/1000~\AA, similar to the slope of 3I/ATLAS. 

A wide range of spectral slope measurements were reported for 2I/Borisov, with values as low as 5\%/1000~\AA\ and as high as 22\%/1000~\AA, depending on the measurement method and wavelength coverage \citep[e.g.,][]{Opitom:2019-borisov, Fitzsimmons:2019, Yang2020, Guzik:2020, Hui2020, Kareta2020, deleon2020, Lin2020, Aravind2021, MazzottaEpifani2021, Prodan2024, Deam2025}. While a meta-analysis of the spectral slopes of 2I/Borisov is beyond the scope of this work, we highlight the \citet{Fitzsimmons:2019} measurement of $\sim$20~\%/1000~\AA\ for the wavelength range 3900~\AA~$< \lambda <$~6000~\AA. The similar wavelength range in our spectrum further supports the claim that 3I/ATLAS is slightly redder than 2I/Borisov. 

Overall, 3I/ATLAS appears redder than the previous interstellar objects, likely reflecting compositional and/or particle size differences between it and 1I/\Oumuamua and 2I/Borisov. All three interstellar objects exhibit colors slightly redder than D-type asteroids \citep{Tholen1984, DeMeo2009} but not as red as some outer solar system objects such as 5145/Pholus \citep{Binzel1992}.

It is uncertain if 3I/ATLAS is older than 1I/\Oumuamua or 2I/Borisov, but its higher velocity may suggest this is the case \citep[e.g.,][]{Taylor2025, Hopkins2025b}. Alternatively, \citet{delaFuenteMarcos2025} and \citet{Guo25} propose an origin in the thin disk, but this does not necessarily contradict the possibility of 3I/ATLAS being an old object
\citep{Hopkins2025b}. An older age might correspond to a higher $\mathrm{H}_2\mathrm{O}$ mass fraction, \citep{Hopkins2025b}, but this may contradict the \emph{JWST} data, which showed about the same CO relative to $\mathrm{H}_2\mathrm{O}$ as 2I/Borisov and much higher $\mathrm{CO}_2$ (never directly measured in 2I/Borisov, see, e.g., \citealp{McKay24}) relative to $\mathrm{H}_2\mathrm{O}$ compared to almost all solar system comets.

There is an emerging debate in the literature about the extent to which 3I/ATLAS has been altered by space weathering. \citet{Yaginuma25} present thermal models that suggest 3I/ATLAS preserved a large fraction of its original volatile material; however, \citet{Maggiolo25} claim significant galactic cosmic ray processing. In any case, space weathering of small bodies in the solar system may redden their surfaces over time \citep[see, e.g.,][]{Brunetto15} and potentially alter their nuclear composition \citep[e.g.,][]{Gronoff20, Maggiolo20}. Some space weathering processes and their relative importance may differ between the solar system and interstellar space, but it seems plausible that the general trend toward reddening over time would be the same. Whether these processes ``saturate'' on a timescale shorter than billions of years remains an open question; if not, 3I/ATLAS's redder spectrum relative to 1I/\Oumuamua and 2I/Borisov could reflect its potentially longer interstellar travel time.

\section{Comments on the Pre-Perihelion Evolution of 3I/ATLAS}

Our finding of minimal spectral slope evolution between July and September 2025, with spectral slopes ranging from 20–29\%/1000~\AA\ and stable $g-r$ colors of 0.69--0.75~mag, may initially appear to contradict reports of dramatic color changes in 3I/ATLAS \citep[e.g.,][]{Seligman2025, Tonry2025}. However, the temporal coverage and activity suggest a two-phase evolutionary model that may reconcile these observations. The first phase, from May ($r_h\approx5.8$~au) to June ($r_h\approx3.9$~au) 2025, is a pre-discovery color transition that entails 3I/ATLAS evolving from a plateau into a brightening phase, followed by the second phase of color stability after cometary activity began.

\subsection{Precovery Plateau Through Brightening Phase}

During the precovery plateau phase (May 15–24, 2025, $r_h\approx5.8$--5.6~au), the dust-to-nucleus flux ratio remained relatively constant, suggesting steady-state dust production with minimal evolution \citep{Ye2025}. ZTF color measurements during this period were consistent with near-solar colors \citep{Seligman2025}. However, \citet{Ye2025} note that the nucleus contribution represents at most $\sim$17\% of the total flux even during this plateau phase, indicating that the object was already actively producing dust. The true nuclear absolute magnitude is constrained at $H_{V}\gtrsim15.4$~mag from HST observations \citep{Jewitt2025}, significantly fainter than the $H_V\approx12.4$~mag measured by ZTF during May, confirming substantial dust contamination even during the apparent color ``plateau'' phase. 

Shortly thereafter (May 27–July 20, 2025, $r_h\approx5.5$--3.9~au), a brightening phase began during which the dust-to-nucleus flux ratio linearly increased from at least $F_\mathrm{dust}/F_\mathrm{nucleus}\approx5$ to $F_\mathrm{dust}/F_\mathrm{nucleus}\approx15$ over $\sim$50 days \citep{Ye2025}. This uniform brightening phase is characterized by a steep heliocentric distance dependence of $\propto r_h^{-3.8}$, significantly steeper than 2I/Borisov ($\propto r_h^{-2.1}$) and more consistent with dynamically old Solar System comets \citep{Ye2025}. Rubin Observatory/LSST precovery observations (from June 21, 2025, at $r_h = 4.85$~au) provide further early-time evidence of detected cometary activity with a clearly resolved dust coma \citep{Chandler2025}.

The dramatic color reddening observed by \citet{Seligman2025} occurred during this uniform brightening phase. However, \citet{Puzia2025} caution that the opposition effect at such low phase angles could partially or entirely explain the apparent color evolution between May and June \citep[see, e.g.,][]{Rosenbush2009}, complicating the interpretation of the early color measurements. By the time of discovery, observations showed clear evidence of reddening and a resolved coma; the opposition effect alone cannot account for the observed red colors \citep{Puzia2025, Opitom2025, Seligman2025}. Based on combined ZTF \citep{Ye2025} and TESS \citep{Feinstein2025, Martinez-Palomera2025} precovery observations, \citet{Ye2025} infer that constant dust outflow began $\sim$30~days before the precovery detections in early April at a distance of $r_h\approx7.5$~au\footnote{We correct the original \citet{Ye2025} manuscript distance, which was erroneously stated as 9~au.}. 

\subsection{Stable Color Phase}
Our spectral time series, spanning from July 3, 2025 ($r_h\approx4.2$~au) to September 2, 2025 ($r\approx2.5$~au), samples the object after most of the major color transition has completed. Throughout this 2-month period, we observe consistently red spectral slopes (20–29\%/1000~\AA) and stable colors ($g-r = 0.69$--0.75~mag, $r-i = 0.26$--0.30~mag, $c-o = 0.50$--0.55~mag), excluding the July 12--14 data.

This color stability persists despite clear evidence of evolving cometary activity during the spectral time series. Water activity was first marginally detected via OH emission at 3085~\AA\ by \citet{Xing2025} using \emph{Swift}/UVOT on July 31 at $r_h\approx3.5$~au. Their derived water production rate implies an active area of at least 7.8~km$^2$. Such an area, if confined to the nucleus surface, would require more than 8\% of it to be active, a larger fraction than observed for most Solar System comets \citep{Xing2025}. Near-infrared spectroscopy revealed the presence of icy grains in the coma, potentially serving as an extended source of water vapor \citep{Yang2025, Xing2025}.

CN emission was absent in early July spectroscopy \citep[][]{Seligman2025, Puzia2025, delaFuenteMarcos2025}, with the first detections of CN in mid-August \citep{SalazarManzano2025, Rahatgaonkar2025, Hoogendam25_KCWI}, preceded by emerging Ni emission features \citep{Rahatgaonkar2025, Hoogendam25_KCWI} and followed by emerging Fe emission features \citep{Hutsemekers25}. JWST NIRSpec observations revealed a $\mathrm{CO_2}$-dominated coma at $r_h = 3.3$~au, along with detections of $\mathrm{H_2O}$, CO, OCS, water ice absorption, and dust continuum \citep{Cordiner2025, Yang2025}. The measured $\mathrm{CO_2/H_2O}$ ratio is larger than the general trend observed in Solar System comets at $r_h\approx3$~au, strongly suggesting $\mathrm{CO_2}$-driven activity. Dust production rates also increased during this period \citep{SalazarManzano2025}. Despite these changes in activity, the spectral slope remained stable throughout August and into September in our observations.

\subsection{A Potential Physical Explanation}

The two-stage evolution (plateau followed by uniform brightening) suggests distinct physical regimes: The plateau phase (May 15–24, 2025, $r_h\approx5.8–5.6$~au) likely represents an initial phase in which constant dust production from the nucleus balanced the geometric dilution of dust particles expanding radially at constant velocity \citep{Jewitt87}. The near-solar colors measured during this phase require careful interpretation, given: (1) dust already dominated the brightness \citep{Ye2025}, (2) the measurements were obtained at low phase angles where opposition effects can significantly brighten and alter colors \citep{Rosenbush2009, Puzia2025}, and (3) uncertainties may be underestimated during the plateau phase due to crowded-field photometry complications \citep[e.g.,][]{Ye2025}.

One potential explanation is that 3I/ATLAS exhibited continuous dust production, potentially dating back as far as $r_h\approx7.5$~au, (early April 2025) \citep{Ye2025}, rather than a sudden transition from an inactive nucleus to a dusty coma. Activity could have potentially started even earlier than $r_h\approx7.5$~au: JWST detected $\mathrm{CO}$ and $\mathrm{CO}_2$ at $r_h=3.32$~au \citep{Cordiner2025}, and both species sublimate further out than $\sim8$~au. The surface sublimation of $\mathrm{CO}_2$ theoretically begins at $\sim$17--20~au and CO sublimation beyond the Kuiper belt for solar system comets. 

The uniform brightening marks a transition where the dust production in 3I/ATLAS increases. The steep brightening law ($\propto r_h^{-3.8}$) is characteristic of dynamically old Solar System comets, consistent with the inferred old age of 3–11~Gyr \citep{Hopkins2025b, Taylor2025, Jewitt25_preperihelion}. The reddening occurred as dust production increased, with the red color likely arising from an increasingly dominant refractory organic component in the larger dust population \citep[see, e.g.,][]{Levasseur-Regourd2018}. These observations suggest that 3I/ATLAS was continuously active from as early as April ($r_h\approx7.5$~au) through discovery. This activity is likely an evolution from low-level to high-level dust production, with associated changes in either dust properties (like grain size) or observational effects (including phase angle) producing the observed color evolution.

After its discovery in early July, 3I/ATLAS maintained a consistently red color despite continued evolution in gas emission and dust production. This can be explained if the spectral slope is primarily determined by the optical properties and size distribution of the entrained dust, which remain relatively constant once a steady-state balance is achieved between dust production and removal. The slope change observed between July 12--14 in our spectral time series may coincide with a brief change in dust properties with large dust grains lifted through micron-sized $\mathrm{H}_2\mathrm{O}$ ice grain sublimation. Outside of the July 12--14 window, the spectral slope during this stage (20–29\%/1000~\AA\ in our measurements) is consistent with typical cometary dust continua \citep{protopapa2018, kareta_noonan23, Kolokolova2024} and slightly redder than D-type asteroids \citep{DeMeo2009}, suggesting a dust population dominated by sub-micron to micron-sized particles with significant refractory organic content.

The increasing CN, Ni, and Fe emission, along with the rising dust production rates observed between August and September \citep{SalazarManzano2025}, do not produce measurable color changes. We speculate these quantities scale together, maintaining a relatively constant dust-to-gas ratio and particle size distribution. \citet{SalazarManzano2025} measured $\log(\mathrm{Q(CN)/Af\rho)}\approx22.4$, consistent with typical gas-to-dust ratios in active comets, suggesting that the dust and gas production are coupled through a common sublimation-driven mechanism.

\section{Conclusions}

We present spectrophotometric data taken using the SNIFS spectrograph on the UH~2.2-meter telescope. Our spectral time series comprises nine spectra spanning the first two months of 3I/ATLAS after discovery, with a wavelength range of 3200 \AA\ to 10000 \AA. Our observations span from shortly after the discovery announcement ($\sim$12.5~hours) to early September, and cover the beginning of optical gas emission activity, contributing to worldwide efforts to characterize 3I/ATLAS. 

Our deepest upper limit is on July 12, with $\mathrm{Q(CN)}< 9.8 \times10^{23}$~molecules~s$^{-1}$. We report a weak detection of CN on August 18 with a production rate of $\left(8.4\pm3.2\right)\times10^{23}$~molecules~s$^{-1}$ using a 3\arcsec\ aperture radius. We subsequently detected CN on August 23 and September 2. Our CN production rate estimates are consistent with other studies \citep[e.g.,][]{Rahatgaonkar2025, Hoogendam25_KCWI, SalazarManzano2025}. We also observe previously reported Ni emission \citep{Rahatgaonkar2025, Hoogendam25_KCWI}. We may barely detect the Fe emission reported in \citet{Hutsemekers25} in our September 2 spectrum, but the detection would be marginal at best. 

The spectral slope of 3I/ATLAS is neutral to red with values from 5\%/1000~\AA\ \citep[see][]{Yang2025} to 39\%/1000~\AA, depending on the measured wavelength. At the bluer wavelengths, e.g., the $g$ band, there is a steep red slope, whereas at redder wavelengths, e.g., in the $z$ band, the slope is nearly flat. This slope is generally redder than reported slopes for 1I/\Oumuamua and 2I/Borisov. 

Comparing the spectral properties of the interstellar comets 1I/\Oumuamua, 2I/Borisov, and 3I/ATLAS to those of analogous small bodies in our own solar system (active and dormant comets, asteroids, Trans-Neptunian Objects, etc.) provides heuristic power to evaluate solar system and planetary formation models. Continued post-perihelion photometric and spectroscopic follow-up will further illuminate this exciting interstellar interloper and provide additional insights into its history, evolution, and composition.

\begin{acknowledgments}

We thank the referee for a thoughtful and thorough review of the manuscript. We thank John Noonan, Theodore Kareta, Daniela Iglesias, Cyrielle Opitom, and Javier Licandro for helpful discussions about 3I/ATLAS and this manuscript. 

W.B.H. acknowledges support from the NSF Graduate Research Fellowship Program under Grant No. 2236415. 

The Shappee group at the University of Hawai\okina i is supported with funds from NSF (grant AST-2407205) and NASA (grants HST-GO-17087, 80NSSC24K0521, 80NSSC24K0490, 80NSSC23K1431).

This work was supported in part by the Director, Office of Science, Office of High Energy Physics of the U.S. Department of Energy under Contract No. DE-AC02-05CH11231

K.J.M., J.J.W., and A.H.\ acknowledge support from the Simons Foundation through SFI-PD-Pivot Mentor-00009672. 

J.T.H. acknowledges support from NASA through the NASA Hubble Fellowship grant HST-HF2-51577.001-A, awarded by STScI. STScI is operated by the Association of Universities for Research in Astronomy, Incorporated, under NASA contract NAS5-26555.

\end{acknowledgments}

\bibliography{3I-SNIFS}{}
\bibliographystyle{aasjournal}

\end{document}